\documentclass[sigconf]{acmart}

\usepackage{adjustbox}
\usepackage{balance} 
\usepackage{booktabs, caption, multirow, makecell}
\usepackage[multiple,para]{footmisc} 
\usepackage{graphics}
\usepackage{subfig}
\usepackage[flushleft]{threeparttable}

\renewcommand{\textrightarrow}{$\rightarrow$}

\copyrightyear{2018}
\acmYear{2018}
\setcopyright{acmcopyright}
\acmConference[DH'18]{2018 International Digital Health Conference}{April 23--26, 2018}{Lyon, France}
\acmBooktitle{DH'18: 2018 International Digital Health Conference, April 23--26, 2018, Lyon, France}
\acmPrice{15.00}
\acmDOI{10.1145/3194658.3194662}
\acmISBN{978-1-4503-6493-5/18/04}

\begin{document}

\title{Eat \& Tell: A Randomized Trial of Random-Loss Incentive to\\ Increase Dietary Self-Tracking Compliance}


\author{Palakorn Achananuparp}
\affiliation{%
  \institution{Singapore Management University}
  \country{Singapore} 
}
\email{palakorna@smu.edu.sg}

\author{Ee-Peng Lim}
\affiliation{%
  \institution{Singapore Management University}
  \country{Singapore} 
}
\email{eplim@smu.edu.sg}

\author{Vibhanshu Abhishek}
\affiliation{%
  \institution{Carnegie Mellon University}
  \city{Pittsburgh} 
  \state{PA} 
}
\email{vibs@andrew.cmu.edu}

\author{Tianjiao Yun}
\affiliation{%
  \institution{Singapore Management University}
  \country{Singapore} 
}
\email{pepperyun@smu.edu.sg}

\renewcommand{\shortauthors}{Achananuparp et al.}

\begin{abstract}


A growing body of evidence has shown that incorporating behavioral economics principles into the design of financial incentive programs helps improve their cost-effectiveness, promote individuals' short-term engagement, and increase compliance in health behavior interventions. Yet, their effects on long-term engagement have not been fully examined. In study designs where repeated administration of incentives is required to ensure the regularity of behaviors, the effectiveness of subsequent incentives may decrease as a result of the law of diminishing marginal utility. In this paper, we introduce random-loss incentive -- a new financial incentive based on loss aversion and unpredictability principles -- to address the problem of individuals' growing insensitivity to repeated interventions over time. We evaluate the new incentive design by conducting a randomized controlled trial to measure the influences of random losses on participants' dietary self-tracking and self-reporting compliance using a mobile web application called Eat \& Tell. The results show that random losses are significantly more effective than fixed losses in encouraging long-term engagement.

\end{abstract}

%
%

\ccsdesc[500]{Applied computing~Health informatics}
\ccsdesc[300]{Applied computing~Consumer health}

\keywords{Health; Quantified Self; Food Logging; Incentives; Loss Aversion; Unpredictability; Randomized Controlled Trial}

\maketitle

\section{Introduction}

Persistent collection of individuals' lifestyle and behavior data is instrumental in realizing precision medicine's visions of providing treatments and healthcare specifically tailored to individuals \cite{Collins2015}.  Recent advances in mobile sensors, wearable technologies, and applications (also known as apps) have greatly facilitated active and passive self-tracking of a wide range of individual behaviors, including physical activities, sleep, and dietary intake. Despite the increasing ease of tracking one own's data, continued engagement with self-tracking technology is not likely to persist long term unless the issues of individual motivation, incentives, and habit formation are also addressed. Recent findings \cite{Clawson2015, Cordeiro2015a} have revealed that a vast number self-tracking technology adopters eventually lost their interest in self-tracking over time, leading to reduced compliance and increased abandonment rate. Many users tended to use the self-tracking data for short-term goals and migration between tools was fairly common \cite{rooksbyetal14chi}.

To promote continued engagement with self-tracking health technologies, various financial and non-financial incentive designs have been proposed. For example, the designs of self-tracking applications can be improved by providing more meaningful behavioral insights from the data \cite{Clawson2015, Cordeiro2015a} and improving social sharing to support observational learning \cite{Clawson2015}. In the context of randomized health intervention trials, financial incentives have been commonly used to motivate health behavior changes and promote self-tracking compliance \cite{Patel2016, Finkelstein2016}. Several researchers \cite{Volpp2008, Volpp2009, Charness2009, Gine2009} argued that financial incentive is a strong motivator for improving many health-related behaviors with little to no crowding-out effect \cite{Charness2009, Gneezy2011, Promberger2013}. However, a few studies \cite{Volpp2009, Finkelstein2016} have also shown that achieving effective results comes with a heavy cost. In the past few years, there is a growing body of evidence supporting the use of behavioral economics principles, such as loss aversion \cite{Kahneman1979}, in the design of \textit{cost-effective} incentive programs \cite{Volpp2008, Milkman2014, Patel2016}. Particularly, financial incentives framed as loss have been proven to be a stronger motivator of health behaviors than those typically framed as gain \cite{Volpp2008, Patel2016} (henceforth loss-framed and gain-framed incentives, respectively). This is explained by the fact that people are more likely to act to avoid loss than to acquire gain as the pain of losing is psychologically \textit{twice} as powerful as the pleasure of gaining \cite{Kahneman1979}. In many longitudinal health studies where participants are expected to be repeatedly administered interventions, e.g., being compensated on a daily basis over a long period of time, the decline in intervention effectiveness may occur due to the law of diminishing marginal utility, leading to reduced compliance. The cost-effective design of repeated financial incentives has not been fully explored, especially in the context of loss-framed incentives. Furthermore, data about individuals' day-to-day dietary choices and habits -- captured via self-tracking and self-reporting -- are challenging to acquire even in non precision medicine cases. The demanding nature of dietary data collection underscores the importance of incentive programs necessary for sustaining individual engagement. To our knowledge, few studies have directly examined the effectiveness of financial incentives in increasing individuals' engagement with dietary self-tracking and self-reporting tasks.

Given the research gaps, we propose a new financial incentive design called \textbf{random-loss incentive} to address the issue of declining effectiveness of repeated financial incentives. The proposed incentive incorporates the \textit{unpredictability} principle, commonly used in Gamification \cite{Chou2015} and behavioral reinforcement \cite{Ferster1957}, into the design of a loss-framed incentive. Specifically, we hypothesize that random losses are more effective than fixed losses in promoting sustained engagement and reducing insensitivity to repeated interventions as people tend to be overpessimistic about their chance of suffering higher-than-expected losses. 

We conducted a randomized controlled trial to investigate the effectiveness of the random-loss incentive to improve individuals' engagement with dietary self-tracking and self-reporting via a new mobile web application, developed as part of our research, called the \textbf{Eat \& Tell} app. The app is specifically designed to facilitate the collection of dietary self-tracking and self-reported data through the food diary data import and the self-report survey components. It is also randomized trial friendly. Potential participants were asked to contribute their dietary self-reported data over a period of 30 days. Upon the successful enrollment, participants were randomly assigned to either the treatment group (random-loss incentive) or the control group (fixed-loss incentive). At the start of the study, they were given an initial S\$35 worth of credit through the app, redeemable for a cash payout after 30 days. The final credit value was determined by their level of compliance with the study protocol. In the treatment condition, a random amount between S\$0 to S\$3 was deducted from the credit balance each day if participants failed to report their dietary intake and behavior on time. On the other hand, control participants were subjected to a fixed S\$1 deduction amount if the same daily requirements had not been met. The sum of all deductions was controlled to never exceed S\$35 in both conditions.

The main contributions of our work are: (1) presenting the Eat \& Tell app as a unified research platform to study individuals' day-to-day dietary habits by leveraging the quantified-self ecosystems of food logging apps, integrated self-report surveys, and incentive programs; (2) introducing a comprehensive set of one-time self-report surveys to collect individual data, e.g., demographic attributes and personality measures, as well as recurring self-report surveys to collect various meal-specific contexts and daily reflection; and (3) introducing a new random-loss incentive that has been shown via a randomized trail of 245 participants to increase the compliance of dietary self-tracking and self-reporting over the fixed-loss incentive. 



\section{Related work}
\label{sec:related}

\subsection{Lifestyle Data and The Quantified Self}
Our study is generally related to the quantified self movement \cite{Swan2013}, an emerging practice of collecting and analyzing one's own data using a wide variety of tools and technologies, including but not limited to modern wearable devices and mobile sensors, to gain a better understanding of and improve certain aspects of daily life. Recently, several studies have explored the use of smartphone, mobile sensing, and wearable technologies to monitor individuals' physical activities and well-being. For instance, Fernandez-Luque et al. \cite{Fernandez-Luque2017} examined the feasibility of capturing the quantified-self data from wearable devices and mobile applications in a weight loss camp for overweight children in Qatar. Kato-Lin et al. \cite{Kato-Lin2016} evaluated the effectiveness of visual-based dietary tracking mobile apps in healthy eating intervention. Rahman et al. \cite{Rahman2016} proposed a sensing framework based on signal processing and machine learning to predict about-to-eat moments for just-in-time interventions. Wang et al. \cite{Wang2014} presented the StudentLife project which employed a variety of smartphone-based mobile sensing techniques to passively and actively track well-being and academic performance of college students. Another related body of work focuses on analyzing self-tracking data in conjunction with other data sources, e.g., social media messages, to learn about individuals' health-related behaviors. For example, De Choudhury et al. \cite{DeChoudhury2017} proposed computational methods to predict individuals' diet compliance success using linguistic, activity, and social capital features extracted from their Twitter messages. Wang et al. \cite{Wang2016} studied weight updates automatically shared on Twitter from a Withings smart scale. In addition, social media content alone has been shown to be a useful data source for a population-level dietary lifestyle monitoring \cite{abbaretal15chi, Mejova2015}. Unlike recent digital health research which utilized observation data to study health-related behaviors, the results of our randomized trial are less susceptible to confounding factors.


A common issue in many health monitoring and intervention studies is the decline in self-tracking compliance over time \cite{Fernandez-Luque2017, Kato-Lin2016, Wang2014}. One reason is that such self-tracking incentive was deliberately omitted to avoid its confounding effects on the intervention outcomes \cite{Kato-Lin2016, Wang2014}. Additionally, study participants are likely to experience tracking fatigue due to the demanding nature of the tasks, resulting in reduced compliance \cite{Cordeiro2015a}. Unlike previous self-tracking studies which focus primarily on health outcomes, such as weight loss, we aim to directly address the declining compliance problem through loss-framed financial incentives. The primary goal of our study is to investigate the effectiveness of different loss-framed incentives in improving self-tracking and reporting compliance in a randomized trial. Compared to past self-tracking studies \cite{Wang2014, Kato-Lin2016, Fernandez-Luque2017}, our Eat \& Tell app is the first to demonstrate an integrative approach to link self-tracking data with context-specific self-report surveys via the use of a unified research and data collection platform, leading to improved user experience. 




\begin{figure*}[thbp]
\centering
\scalebox{0.72}{
\includegraphics[width=\textwidth]{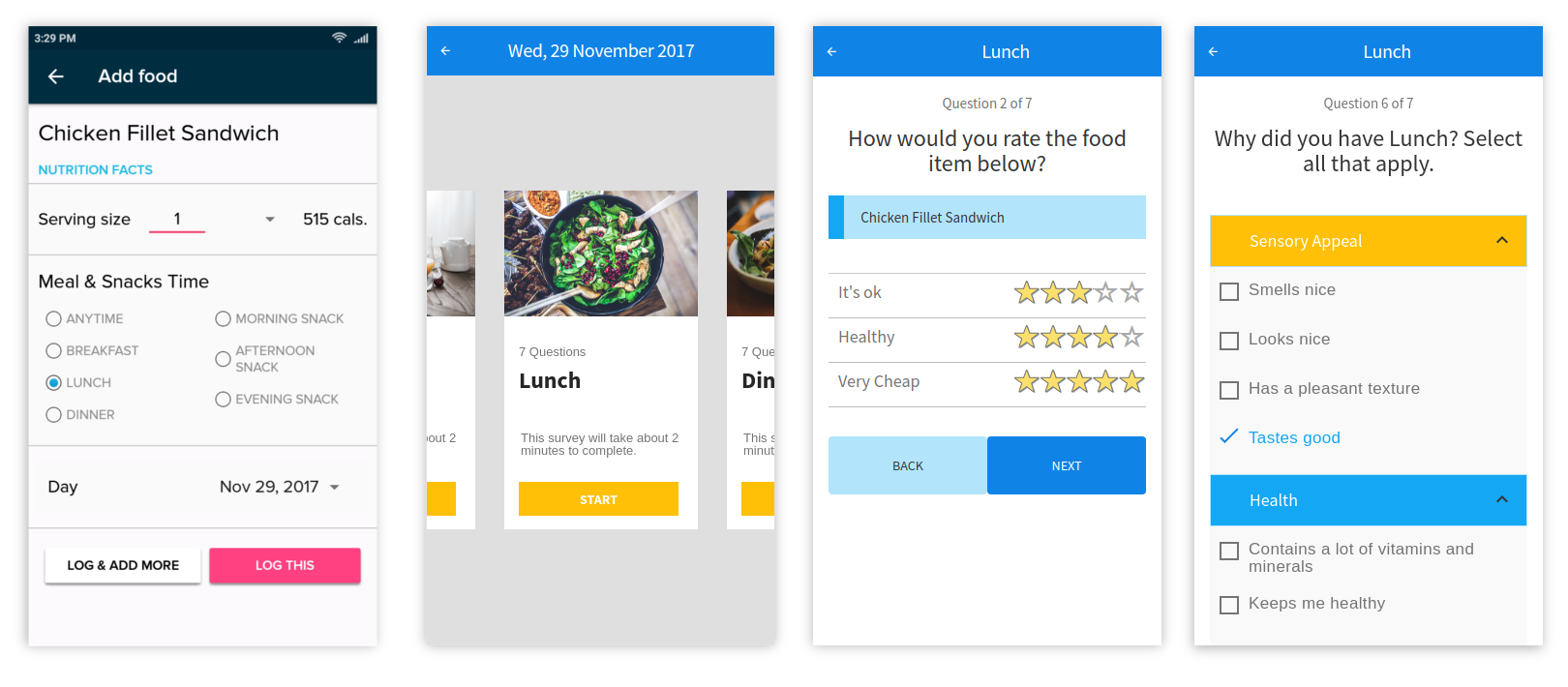}
}
\caption{An example of data integration flow in the Eat \& Tell app}
\label{fig:eatntell}
\end{figure*}

\subsection{Incentives for Health Behavior Change}
Past studies have shown strong evidence of financial incentives in promoting changes in health behaviors, such as smoking cessation \cite{Volpp2009, Gine2009}, weight loss \cite{Volpp2008}, exercising \cite{Charness2009}, and physical activity \cite{Patel2016, Finkelstein2016, Finkelstein2008}. Amongst these studies, behavioral economics principles, such as loss aversion \cite{Kahneman1979} and commitment devices \cite{Bryan2010, Milkman2014}, have been incorporated into the incentive designs to further nudge individuals toward positive behavior changes. For example, Volpp et al. \cite{Volpp2008} showed that individuals who were administered deposit contract and lottery incentives significantly lost more weights than non-incentivized individuals. In a recent study by Patel et al. \cite{Patel2016}, loss-framed financial incentives  were more effective in promoting physical activity than gain-framed incentives. Milkman et al. \cite{Milkman2014} introduced a novel commitment device called temptation bundling which was shown to be effective in increasing gym attendance.

The design of our random-loss incentive is built upon the application of behavioral economics principles, specifically loss aversion and framing, in influencing health-related behavior change \cite{Patel2016}. To our knowledge, this study is the first randomized trial to examine the impacts of unpredictability in loss-framed financial incentives on the individuals' continued engagement with dietary self-tracking and self-reporting. Prior work \cite{Volpp2008, Halpern2011} has utilized unpredictability-based designs in gain-framed incentives (e.g., lottery incentives) to exploit individuals' tendency to overestimate small probabilities of large potential rewards \cite{Kahneman1979}, albeit with mixed results \cite{Halpern2011}. Unlike lottery incentives where the likelihood of a payout is randomly determined, we opt for a fixed reward scheduling and introduce unpredictability during the payout stage similar to Random Rewards in Gamification \cite{Chou2015}, i.e., a randomly determined deduction amount is made for every non-compliance behavior. Lastly, our study is the first to comprehensively evaluate the influences of the incentives in dietary self-tracking studies by controlling for various personal characteristics including self-efficacy \cite{Kato-Lin2016}, grit \cite{Duckworth}, age, and gender, which intrinsically determine individuals' motivation and engagement levels.


\section{Eat \& Tell App}
\label{sec:app}
We begin by describing the Eat \& Tell app\footnote{\url{https://eatntell.sg}}, a mobile web application built by the authors as a research platform for studying daily eating habits and behavior change intervention. To facilitate the data collection process and taking advantage of the quantified self and self-tracking ecosystems, the app integrates data from popular mobile food diary apps to generate personalized self-report surveys for each user based on their food diary records. These surveys are instrumental in capturing additional behavioral and contextual information beyond what is provided by the self-tracking data. For instance, did they have breakfast alone or with other people? Did they go to a restaurant for lunch? Did they choose to have a particular meal for dinner because of sensory appeals or convenience?


\subsection{Food Diary Data Integration}\label{subsec:food_diary}
Food diary data from 3 mobile health and fitness apps, i.e, MyFitnessPal, Fitbit, and Healthy 365\footnote{\url{https://www.healthhub.sg/apps/25/healthy365}} can be automatically synchronized to the Eat \& Tell app via web-based application programming interfaces (APIs). These mobile apps were chosen based on their popularity amongst users in Singapore where the study took place. MyFitnessPal is a popular mobile calorie counter which allows users to track their caloric balance through food and exercise diaries. Next, Fitbit is a well-known commercial service helping its users to track their physical activities, sleep, and food intakes via its activity trackers and mobile app. Lastly, Healthy 365 is a free-to-use health and diet tracking mobile app created by the Health Promotion Board of Singapore. Upon successful data synchronization, a set of self-report surveys are dynamically generated for each user based on the predefined question templates and their food diary data. Each survey is associated with meal of the day (i.e., breakfast, lunch, dinner, and snack) and contains a set of questions about meal contexts (e.g., venue, meal time, eating companions, etc.), food choices, and food items logged in the meal. Figure \ref{fig:eatntell} shows an example of the data integration flow starting from (1) a user logs what he/she had for lunch in the Fitbit app, (2) the Lunch survey is generated in the Eat \& Tell app from his/her Fitbit data, and (3) the user responds to the questions about the food items and (4) the meal itself.

\subsection{Self-Report Surveys}
We administer two major types of self-report surveys in the Eat \& Tell app: \textbf{One-time surveys} and \textbf{daily surveys}. Firstly, one-time surveys are intended to be completed once and consist of standard questionnaires about demographic information (e.g., gender, age, household income level), general health status (e.g., weight, height, blood pressure), physical activity level, and personality measures. To assess the physical activity level, we adopt questions from the Rapid Assessment of Physical Activity (RAPA) \cite{Topolski2006} for their brevity. Next, two personality measures, personality traits and grit, are administered by 50-item IPIP Big-Five Factor Markers \cite{Goldberg1999} and 12-item Grit Scale \cite{Duckworth}, respectively. All users have access to the same set of one-time surveys.

Secondly, daily surveys are intended to be filled out on a daily basis and comprise: (1) a set of dynamically generated surveys (also known as food diary surveys) based on the connected food diary data; and (2) an end-of-day reflection survey. Questions about individual food items and meals of the day are included in the food diary surveys. At a food-item level, we ask users to rate each item in their diary on a 5-point Likert scale about taste (1 = hated it to 5 = loved it), healthfulness (1 = very unhealthy to 5 = very healthy), and affordability (1 = very expensive to 5 = very cheap). To lessen the self-reporting burden, our app keeps track of the users' most frequently-logged items and assign default ratings to similar items. Next, at a meal level, users provide: (1) responses to questions about meal contexts, such as time of the day, type of venue/meal location, eating companions, etc.; and (2) reasons for which a particular meal was chosen in the Food Choice questionnaire \cite{Steptoe1995}. Lastly, the end-of-day surveys include self-assessment questions about the users' subjective well-being and perceived difficulty (self-efficacy) in logging food diary and completing surveys, among others. The subjective well-being questions are adopted from a 12-item Scale of Positive and Negative Experience (SPANE) \cite{Diener2010}, whereas the 5-point scale self-efficacy questions are adapted from the similar self-efficacy questionnaires used in \cite{Kato-Lin2016}. The survey template containing all items used in this study is accessible online\footnote{\url{http://bit.ly/2zHcD9X}}.


\section{Random-Loss Incentive}
\label{sec:incentive}
In this section, we describe random-loss incentive, a new loss-framed financial incentive program designed to address the issue of declining incentive efficacy caused by repeated administration of financial incentives over time. In certain longitudinal-study designs where participants are compensated on a regular basis, such as getting paid each day for achieving the goal of 10,000 steps of physical activity a day, the effectiveness of subsequent payouts of the same amount may be decreasing according to the law of diminishing marginal utility. At the start of the study, a participant may be significantly motivated to act in exchange for the first \$5 payout. However, as the participant's financial gain is accumulating over the course of the study, the next \$5 payout has a proportionately smaller contribution to the total gain and his/her marginal utility. Before intrinsic motivation and habit are properly developed, the same fixed-value payout may have less effect on his/her behavior than past payouts. Similar effects may also apply to loss-framed incentives. In addition to the use of standard behavioral economics principles \cite{Kahneman1979}, including loss aversion, the endowment effect, and framing, the design of our random-loss incentive is also inspired by Random Rewards in Gamification \cite{Chou2015} which is based on the principles of unpredictability and curiosity. The underlying assumption of our random-loss incentive is that people tend to be overpessimistic and emotional about their chance of suffering higher-than-expected losses and are more likely to act to avoid further misfortune than when facing predictable losses.

\begin{figure}[thbp]
\centering
\scalebox{0.6}{
\includegraphics{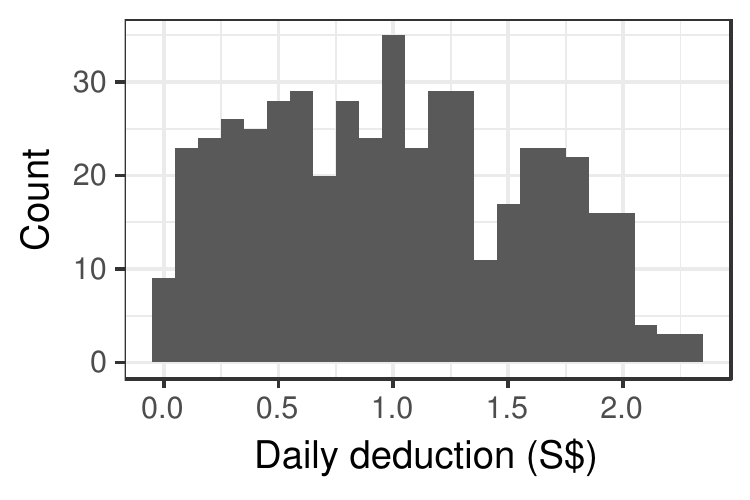}
}
\caption{Distribution of daily deductions.}
\label{fig:hist_random_loss}
\end{figure}

Let $m$ be the length of study ($m$=30 days) and $p$ be the total daily deduction budget ($p$=S\$30), the steps to assign random losses for each treatment participant are as follows. First, we generate a probability vector $l$ of $m$ entries where each entry $l_i$ is a random positive real number between 0 and 1 and the sum of all entries is one. Next, each random loss is computed by multiplying each vector entry $l_i$ by $p$. Finally, we derive for the participant a sequence of $m$ random losses which add up S\$30. This method results in a non-uniform distribution of random losses which are biased toward less extreme values, making them characteristically different from lottery-based incentives. The histogram in Figure \ref{fig:hist_random_loss} shows the distribution of daily deduction amounts (SD = 0.5838) pre-generated under the random-loss incentive program for all treatment participants at the start of the study. As we can see, all random-loss based daily deductions are between S\$0 and S\$3.

\section{Study Design and Procedures}
\label{sec:procedures}

\subsection{Recruitment}
The study was approved by the Institutional Review Board at Singapore Management University. In total, 245 participants were recruited online through the Facebook advertising platform. An advertisement campaign was launched to target Facebook users in Singapore who were 18 years or older, with access to an Android or iOS smartphone, and generally interested in mobile food diary, self-tracking, and wearable technology. To target these sets of Facebook users based on their topics of interests, the following keywords were used: "MyFitnessPal", "Fitbit", "Health Promotion Board, Singapore", "Activity Tracker", "Smartwatch", and "Wearable Technology". Based on these criteria, the campaign approximately reached 400,000 potential audience according to estimates provided by the Facebook platform. The advertisements described an opportunity to participate in a one-month research project to study individuals' daily eating habits and food choices in exchange for a cash compensation of up to S\$35 and a chance to win a smartwatch (approximately S\$500 in value) at the end of the study. Clicking on the advertisements redirected the users to the study website where study details, requirements to participate, and a link to the registration page were shown. The campaign ran for 24 days from May 30 - June 23, 2017. In total, the advertisements were shown 86,533 times (impressions) to 53,635 unique users (reach) yielding 2,934 clicks on the ads or 5.47\% click-through rate (CTR) per unique user. More than 80\% of Facebook users who clicked on the ads were women. Ultimately, 257 website visitors from Facebook signed up to participate yielding 8.76\% conversion rate per website visit at a conversion cost of S\$3.84 campaign spending per sign up. Amongst the potential participants, 5 people failed to complete the registration process, whereas 7 people eventually dropped out a few days into the study by revoking their consent.



\subsection{Procedures}
To register for the study, participants filled out an online registration form on the Eat \& Tell study website. First, they had to give consent to participate, allowing the research team to collect their personal data, online food diary records, and survey responses. Next, they provided their first name, last name, email address, and PayPal account (for receiving compensation) in the registration form. In addition, they were required to connect their online food diary account from one of the supported applications (MyFitnessPal, Fitbit, and Healthy 365) to the Eat \& Tell app, giving the app a permission to access their food diary data. If they had never used an online food diary before, they were allowed to sign up for a new food diary account before resuming the registration process. To complete the registration, participants were required to confirm their identity by clicking on the verification link sent to their registered email address. No multiple sign-ups, e.g., duplicate email addresses or food diary accounts, were allowed. Additionally, we required a smartwatch winner to present a valid national identification card as proof of identity when picking up the prize in person. 

Upon successful registration, participants were randomly assigned to one of the two experimental groups: \textbf{treatment} (random-loss incentive) or \textbf{control} (fixed-loss incentive), after which an introductory email was sent to their registered email address describing the start and end dates of the study and the study procedures. All participants were first instructed to log in to the Eat \& Tell app and fill out a set of 5 one-time surveys about their demographic and health information, personality traits, and grit. Next, for each day in the study, participants were asked to complete a daily task involving (1) logging their food and drink intakes throughout the day in the registered food diary app; (2) synchronizing the food diary data with the Eat \& Tell app and filling out surveys about food choices made in the corresponding diary entries; and (3) filling out a reflection survey about their subjective well-being and perceived task difficulty (self-efficacy) which was accessible after 5:00 PM. In total, it took about 15 - 20 minutes to complete the daily task.

\begin{table}[tbhp]
\centering
\caption{Characteristics of participants.}
\label{tbl:characteristics}
\scalebox{0.8}{
\begin{threeparttable}
\begin{tabular}{@{}lllll@{}}
\toprule
\multicolumn{2}{l}{Variable} & Treatment & Control & All \\ \midrule
\multirow{2}{*}{Gender} & Female & 85 & 66 & 151 \\
 & Male & 31 & 22 & 53 \\ \midrule
\multirow{6}{*}{Age} & 18 - 24 & 19 & 22 & 41 \\
 & 25 - 34 & 55 & 40 & 95 \\
 & 35 - 44 & 21 & 14 & 35 \\
 & 45 - 54 & 6 & 6 & 12 \\
 & 55 - 64 & 3 & 0 & 3 \\
 & 65 or older & 1 & 0 & 1 \\ \midrule
\multirow{4}{*}{Ethnicity} & Chinese & 111 & 83 & 194 \\
 & Indian & 3 & 4 & 7 \\
 & Malay & 1 & 0 & 1 \\
 & Others & 1 & 1 & 2 \\ \midrule
\multirow{5}{*}{Education} & Secondary (O or N level) & 8 & 3 & 11 \\
 & Junior college (A level) & 7 & 5 & 12 \\
 & Vocational certificate & 1 & 4 & 5 \\
 & Polytechnic diploma & 17 & 14 & 31 \\
 & University or higher & 80 & 62 & 142 \\ \midrule
\multirow{7}{*}{Employment} & Full-time employment & 78 & 57 & 135 \\
 & Part-time employment & 4 & 4 & 8 \\
 & Entrepreneur & 1 & 2 & 3 \\
 & Student & 17 & 14 & 31 \\
 & Homemaker & 3 & 5 & 8 \\
 & Unemployed & 3 & 4 & 7 \\
 & Retired & 1 & 0 & 1 \\ \midrule
\multirow{4}{*}{Marital status} & Married & 39 & 27 & 66 \\
 & Living with a partner & 2 & 0 & 2 \\
 & Divorced & 2 & 0 & 2 \\
 & Never been married & 70 & 60 & 130 \\ \midrule
\multicolumn{2}{l}{BMI (mean)} & 23.16 & 22.49 & 22.88 \\ \midrule
\multicolumn{2}{l}{Grit score (mean)} & 2.73 & 2.78 & 2.75 \\ \midrule
\multicolumn{2}{l}{Pre-treatment logging days (mean)} & 1.12 & 1.53 & 1.3 \\ \midrule
\multicolumn{2}{l}{Number of diary entries logged per day (mean)} & 6.23 & 5.81 & 6.06 \\ \midrule
\multicolumn{2}{l}{Caloric intake (kcal) per day (mean)} & 1249.05 & 1263.05 & 1254.85 \\ \midrule
\multicolumn{2}{l}{Number of participants ($N$)} & 135 & 110 & 245 \\ \midrule
\multicolumn{2}{l}{Number of intervention participants ($N'$)} & 111 & 91 & 202 \\ \bottomrule
\end{tabular}
\begin{tablenotes}
\small
\item $N$ also includes participants who did not answer/partially answered one-time survey questions. Intervention participants are those who were successfully enrolled into the study and received at least 1 deduction.
\end{tablenotes}
\end{threeparttable}}
\end{table}

Both treatment and control participants were then informed in the introductory email that a \$35 online credit had been deposited to their Eat \& Tell account. The credit was redeemable for real money of the same value at the end of the study, paid to them via the online payment platform Paypal. However, their online credit was still subjected to being deducted during the course of the study depending on the completion of the one-time surveys and the daily tasks. First, \$1 would be deducted from their credit balance for each incomplete one-time survey for all participants. All one-time surveys were accessible until the end of the study. Next, participants in each group were told about the deduction amount for each incomplete daily task specific to their assigned condition: (1) a random deduction between S\$0 to and S\$3 for the treatment group; and (2) a S\$1 deduction for the control group. All daily tasks had to be completed within the next 48 hours (calendar time). Any overdue surveys were automatically made inaccessible to participants after its deadline had passed. Next, all participants were instructed that the total deduction amount would never exceed S\$35 (S\$5 for one-time surveys and S\$30 for all daily tasks). In addition, their chance of winning a smartwatch was also proportional to their task completion rate. This was done to discourage anyone who signed up just to be in the smartwatch sweepstake but never intended to contribute to the study.

\subsection{Automated Email Notifications}
To facilitate participants' compliance with the study protocol, the Eat \& Tell app sent an automated email reminder at 7:00 PM every day to any participants who had not yet completed the daily task, notifying them about the remaining time and a potential credit loss. Participants were allowed to change the schedule of the email reminder by specifying the preferred time in the app. In total, two email reminders were sent before the daily task's due date. Ultimately, if participants did not complete the daily task on time, they would receive another automated email message at 8:00 AM the next day after the due date, notifying them about the recent deduction and offering short motivational messages and helpful tips on how to make the daily tasks less tedious.

\subsection{Participants}
All 245 participants who successfully enrolled in the study were randomly assigned to one of the two experimental groups: treatment ($N$=135) and control ($N$=110). Participant characteristics are shown in Table \ref{tbl:characteristics}. Despite the S\$5 incentive to filling out one-time surveys, not all participants complied and thus some of the participant demographics and characteristics were missing. A large portion of the study participants were female (74.02\%), which was in line with other studies of the same nature \cite{Kato-Lin2016}. There were no significant differences in the gender distribution between the conditions. Next, 50.8\% of participants were 25 - 34 years old. Almost all participants (95.1\%) were Chinese, which was overly represented compared to the population of Singapore \cite{web:singstat2017}. Next, 70.65\% of participants had a university degree or higher and 69.95\% had a full-time employment. More than half of the study participants (65\%) had never been married. Next, the average body mass index (BMI), calculated from participants' self-reported weight and height, of the study participants was 22.88 (SD = 5.34), which was near the normal weight and overweight boundary for Asian populations. Additionally, the BMI distribution was left-skewed, i.e., there were more normal-weight participants than overweight and obese ones. Next, the average grit score of participants in our study was 2.75 (SD = 0.58), which is considered to be moderate. Lastly, only a small portion of participants ($N$=25; 10.2\%) had regularly logged food diaries prior to joining the study. On average, all participants logged 1.3 days (SD = 4.3; median = 0) of food diary in the last 30 days before enrolling in the study. There were no significant differences in BMI, grit score, and pre-treatment logging experience between the experimental groups.

\begin{figure}[tbp]
\centering
\scalebox{0.6}{
\includegraphics{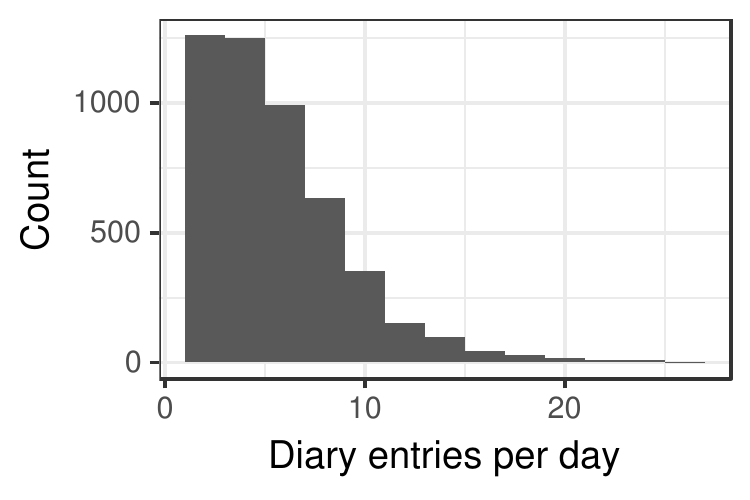}
}
\caption{Distribution of daily diary entries logged.}
\label{fig:hist_entries}
\end{figure}

\begin{figure}[tbp]
\centering
\scalebox{0.6}{
\includegraphics{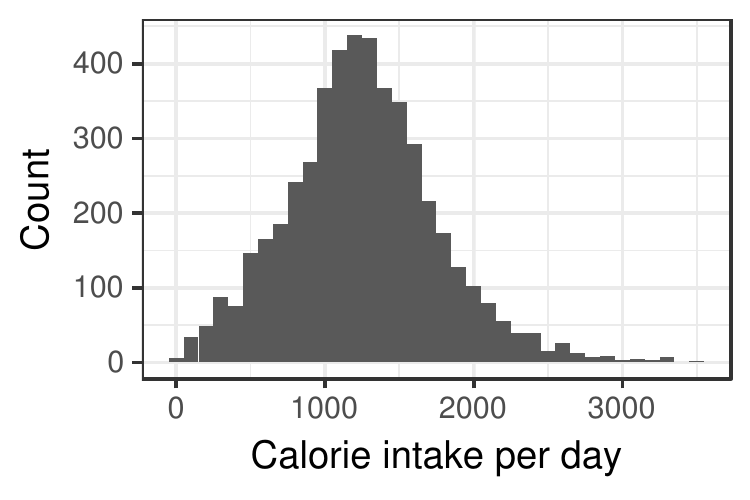}
}
\caption{Distribution of daily caloric intakes (kcal).}
\label{fig:hist_calories}
\end{figure}

Next, we examine the participants' food logging and eating behavior during the study. The distribution of the number of diary entries logged per day is shown in Figure \ref{fig:hist_entries} where the mean is 6.06 entries (SD = 3.64). On average, treatment participants logged 6.23 entries per day (SD = 3.81), whereas control participants logged 5.81 entries per day (SD = 3.38) as shown in Table \ref{tbl:characteristics}. There is a significant difference between the experimental groups according to the Mann-Whitney U test of diary entries ($U$=2704500, $p$<0.01). Next, to examine the participants' daily caloric intakes, we first removed 3 outlying data points whose daily caloric intakes (e.g., more than 100K kcal consumed in a single day) were higher than two standard deviations away from the participants' mean caloric intake. Upon further inspection, we found that the outliers were caused by the inaccuracy in the user-contributed food databases. Figure \ref{fig:hist_calories} displays the distribution of the amount of caloric intake per day. As we can see from Table \ref{tbl:characteristics}, on average, all participants consumed 1254.85 kcal per day (SD = 508.83), whereas treatment and control participants consumed 1249.05 kcal (SD = 520.14) and 1263.05 kcal (SD = 492.4) per day, respectively. Although control participants had a slightly higher mean caloric intake than treatment participants, the difference is not statistically significant.

Lastly, at the end of the study, 202 participants (111 and 91 in the treatment and the control conditions, respectively) were considered intervention participants, i.e., those who received at least 1 deduction and therefore not having full compliance. There is no significant difference in the numbers of intervention participants between the two experimental groups.

\section{Analysis and results}
\label{sec:analysis}
The primary goal of our study is to evaluate the effectiveness of random-loss and fixed-loss incentives by measuring the participants' self-tracking and self-reporting (i.e., daily task) compliance during the 30-day intervention period. Moreover, our secondary goal is to examine the effects of the two incentive programs in promoting sustained behavior change after the inventions are removed. The analysis was conducted on a modified intention-to-treat (mITT) basis. Specifically, we exclude 43 participants (24 from the treatment group and 19 from the control group) who did not receive full interventions from the analysis, i.e., those who were only briefed about the financial incentive program at the start of the study but never received any deductions and related email messages throughout the study due to full compliance. Ultimately, 202 participants are included in the analysis.

\subsection{Preliminary Analysis}
\label{subsec:prelim}
\begin{figure}[thbp]
\centering
\scalebox{0.6}{
\includegraphics{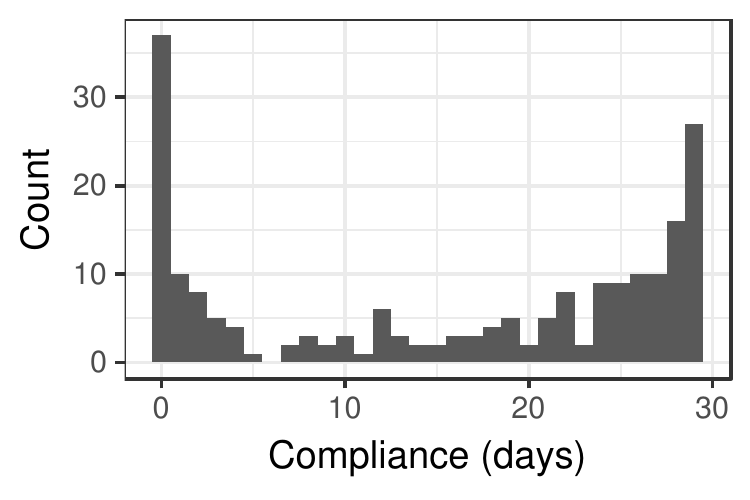}
}
\caption{Distribution of the number of days participants completed daily tasks.}
\label{fig:hist_compliance_days}
\end{figure}

We first present a preliminary analysis to examine the participants' compliance in completing the daily tasks across the experimental groups. Our outcomes of interest are \textbf{compliance days}, defined as the number of days a participant completed his/her daily task, and \textbf{daily compliance users}, defined as the percentage of participants who completed the daily task on a specific day. The distribution of the number of days of completed daily tasks by participants from both treatment and control conditions is shown in Figure \ref{fig:hist_compliance_days}. The median number of days of completed daily tasks is 19. In total, 37 participants (18\%) did not complete any daily tasks.

\begin{figure}[thbp]
\centering
\scalebox{0.6}{
\includegraphics{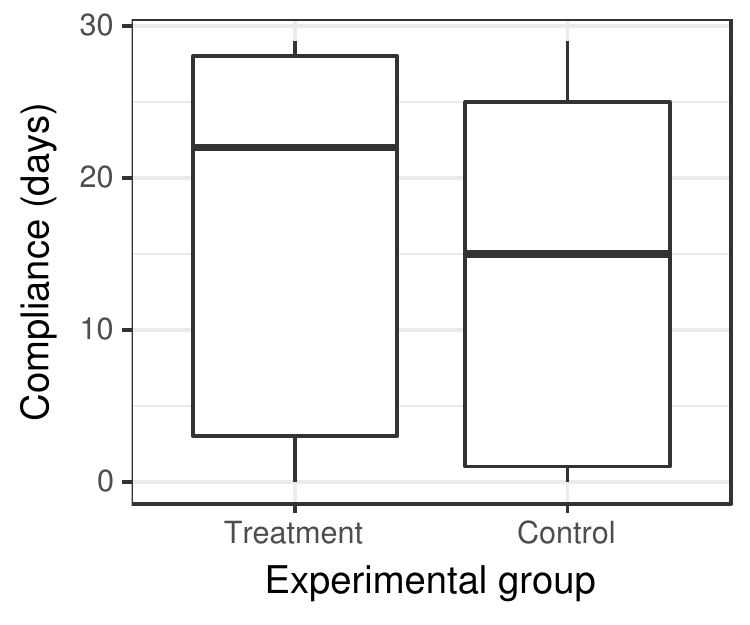}
}
\caption{Compliance days by experimental group.}
\label{fig:boxplot_compliance_days}
\end{figure}

Figure \ref{fig:boxplot_compliance_days} shows a box plot of compliance days for the two experimental groups. As we can see, treatment participants generally have a higher level of compliance with the study protocol than control participants. The medians compliance days of the treatment and the control groups are 22 and 15, respectively. The nonparametric Mann-Whitney U test of compliance days in the two conditions over the 30-day period shows a significant difference ($H_0$: treatment = control; $U$=4142, $p$<0.05). Uncontrolled, the analysis suggests that our random-loss incentive is more effective than the fixed-loss incentive in improving participants' daily task completion over 30 days.



\begin{figure}[thbp]
\centering
\scalebox{1.0}{
\includegraphics[width=\columnwidth]{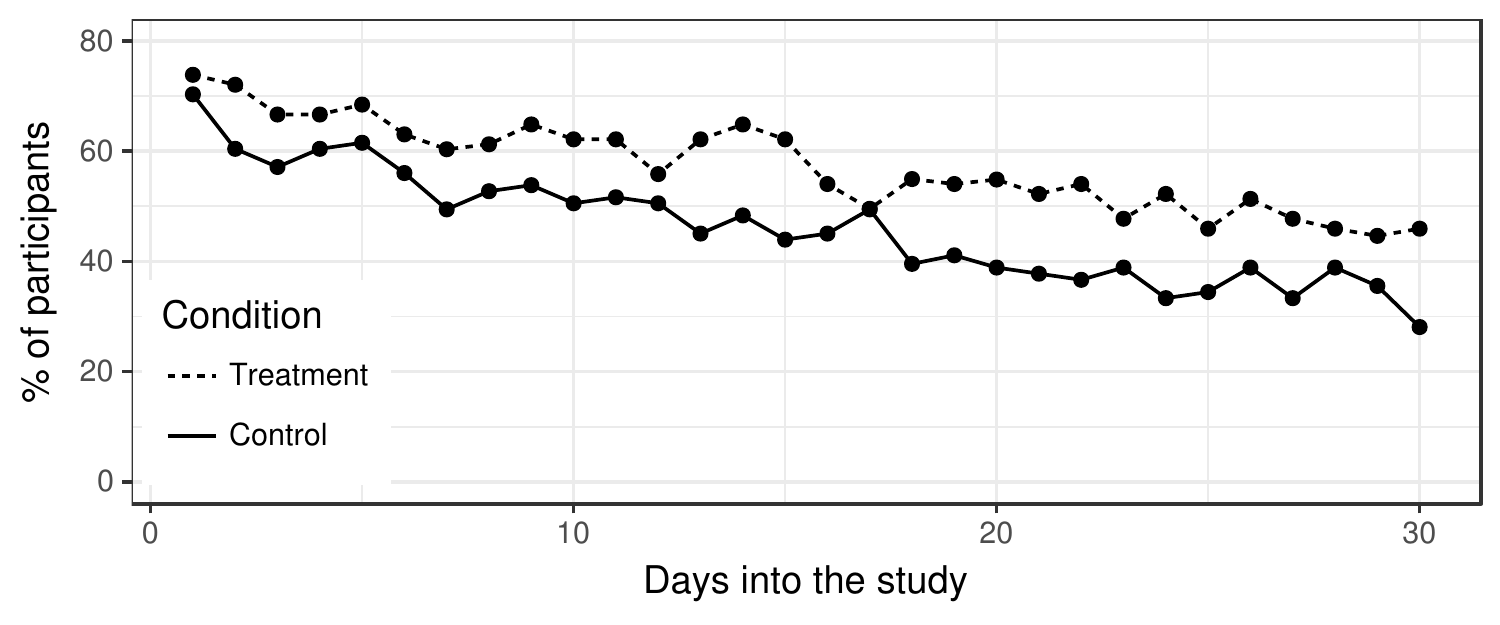}
}
\caption{Daily compliance users over time.}
\label{fig:lines_compliance_days}
\end{figure}

Figure \ref{fig:lines_compliance_days} depicts the percentages of daily compliance users from the start to the end of the study. On the first day, both conditions have the similar daily compliance users (73.87\% and 70.33\% for treatment and control, respectively). However, the compliance trend continues to decrease over time as the study progresses. This is not surprising since the declining in compliance is consistent with past studies on longitudinal health behaviors \cite{Patel2016, Kato-Lin2016, Milkman2014}. Over the 30-day period, the treatment condition has consistently shown higher daily compliance users than the control condition. As we can see, 45.95\% of treatment participants completed the daily task on the last day of the study, whereas only 28.09\% of control participants did so.

\subsection{Effectiveness of the random-loss incentive}
In this section, we present the regression analysis to assess the effectiveness of our random-loss incentive. Due to the repeated measures design of the study, we model participants' daily compliance using mixed effects logistic regression, taking into account that data points from the same participant are not independent. The binary dependent variable in our models is the participants' day-to-day compliance status (1 = a participant completed the task in time on a given day, otherwise 0). Main effects for experimental condition (binary predictor; 1 = treatment, otherwise 0) and interactions between experimental condition and other predictors (e.g., the number of days into the study for a given participant) are included. Next, the models control for each participant's number of days into the study (from 1 to 30). To adjust for individuals' pre-treatment differences in self-tracking experience and self-efficacy, we also control for the number of food-diary logging days in the past 30 days before the study started and self-reported perceived task difficulty for a given day in the study, respectively. Lastly, personal characteristics, such as gender (binary predictor; 1 = male, otherwise 0), age, and self-reported grit score (from 1 to 5), are also controlled for. All analyzes were performed using the lme4 package (version 1.1.14) \cite{Bates2015} in R (version 3.4.2).

\begin{table*}[tbp]
\centering
\caption{Mixed-effects logistic regressions for daily task completion.}
\label{tbl:mixed_effects}
\scalebox{0.86}{
\begin{threeparttable}
\begin{tabular}{@{}lllllllll@{}}
\toprule
Variables & Model 1 & Model 2 & Model 3 & Model 4 & Model 5 & Model 6 & Model 7 & Model 8 \\ \midrule
\textbf{FIXED EFFECTS:} & \multicolumn{8}{l}{\textbf{ESTIMATE}} \\ \midrule
(Intercept) & -0.71485$^*$ & 0.78933$^*$ & 1.00936$^{**}$ & 0.80661$^*$ & 5.77789$^{**}$ & 7.88429$^{**}$ & 5.88135$^*$ & 5.63715 \\
\textbf{Primary predictor variables} &  &  &  &  &  &  &  & \\
\hspace{5mm}Treatment & 0.90950$^*$ & 1.01532$^*$ & 0.60770 & 0.62142 & -1.63751$^\dag$ & -2.45735$^*$ & 0.93464 & -0.70462 \\
\hspace{5mm}Days into the study $\times$ Treatment & - & - & 0.02566$^*$ & 0.02585$^*$ & 0.09089$^{**}$ & 0.08749$^*$ & 0.07013$^*$ & 0.06986$^*$ \\
\hspace{5mm}Male $\times$ Treatment & - & - & - & - & - & 2.84370 & 2.14528 & 2.32768 \\
\hspace{5mm}Age $\times$ Treatment & - & - & - & - & - & - & -0.10341 & -0.09940 \\
\hspace{5mm}Grit score $\times$ Treatment & - & - & - & - & - & - & - & 0.52969 \\
\textbf{Control variables} &  &  &  &  &  &  &  & \\ 
\hspace{5mm}Days into the study & - & -0.10215$^{**}$ & -0.11662$^{**}$ & -0.11674$^{**}$ & -0.16769$^{**}$ & -0.17019$^{**}$ & -0.13508$^{**}$ & -0.13469$^{**}$ \\
\hspace{5mm}Pre-treatment logging days & - & - & - & 0.11290$^*$ & -0.03017 & -0.03697 & -0.04176 & -0.04022 \\
\hspace{5mm}Perceived task difficulty & - & - & - & - & 1.45901$^{**}$ & 0.99555$^\dag$ & 1.12445$^{**}$ & 1.14025$^{**}$ \\
\hspace{5mm}Male & - & - & - & - & - & -2.52521 & -1.30442 & -1.24929 \\
\hspace{5mm}Age & - & - & - & - & - & - & 0.00871 & 0.09400 \\
\hspace{5mm}Grit score & - & - & - & - & - & - & - & 0.05229 \\ \midrule
\textbf{RANDOM EFFECTS:} & \multicolumn{8}{l}{\textbf{VARIANCE}} \\ \midrule
Subjects (intercept) & 8.415 & 10.47 & 10.52 & 10.26 & 9.806 & 12.223 & 3.241 & 3.028 \\
Perceived task difficulty (slope) & - & - & - & - & 3.073 & 2.206 & 1.334 & 1.312 \\ \midrule
\textbf{Model statistics} &  &  &  &  &  &  &  & \\
\hspace{5mm}AIC & 4779.5 & 4354.4 & 4350.8 & 4347.8 & 824.8 & 772.1 & 703 & 706.6 \\
\hspace{5mm}BIC & 4799.6 & 4381.3 & 4384.3 & 4388 & 879.8 & 838.5 & 780.1 & 795.5 \\
\hspace{5mm}Number of observations & 6050 & 6050 & 6050 & 6050 & 3316 & 3114 & 2774 & 2774 \\
\hspace{5mm}Number of subjects & 202 & 202 & 202 & 202 & 168 & 150 & 136 & 136 \\ \bottomrule
\end{tabular}
\begin{tablenotes}
\small
\item Significance thresholds are $\dag$ $p$<0.1, * $p$<0.05, and ** $p$<0.01.
\end{tablenotes}
\end{threeparttable}}
\end{table*}

The results are presented in Table \ref{tbl:mixed_effects}. First, the uncontrolled baseline model (Model 1) shows that, on any given day, treatment participants are significantly more likely ($p$<0.05) to complete the daily tasks compared to control participants. This suggests that the random-loss incentive is generally more effective than the fixed-loss incentive in influencing the participants' engagement. The result is also in line with the Mann-Whitney U test of aggregated participant compliance presented in the preliminary analysis. 

\begin{figure}[thbp]
\centering
\scalebox{1.0}{
\includegraphics[width=\columnwidth]{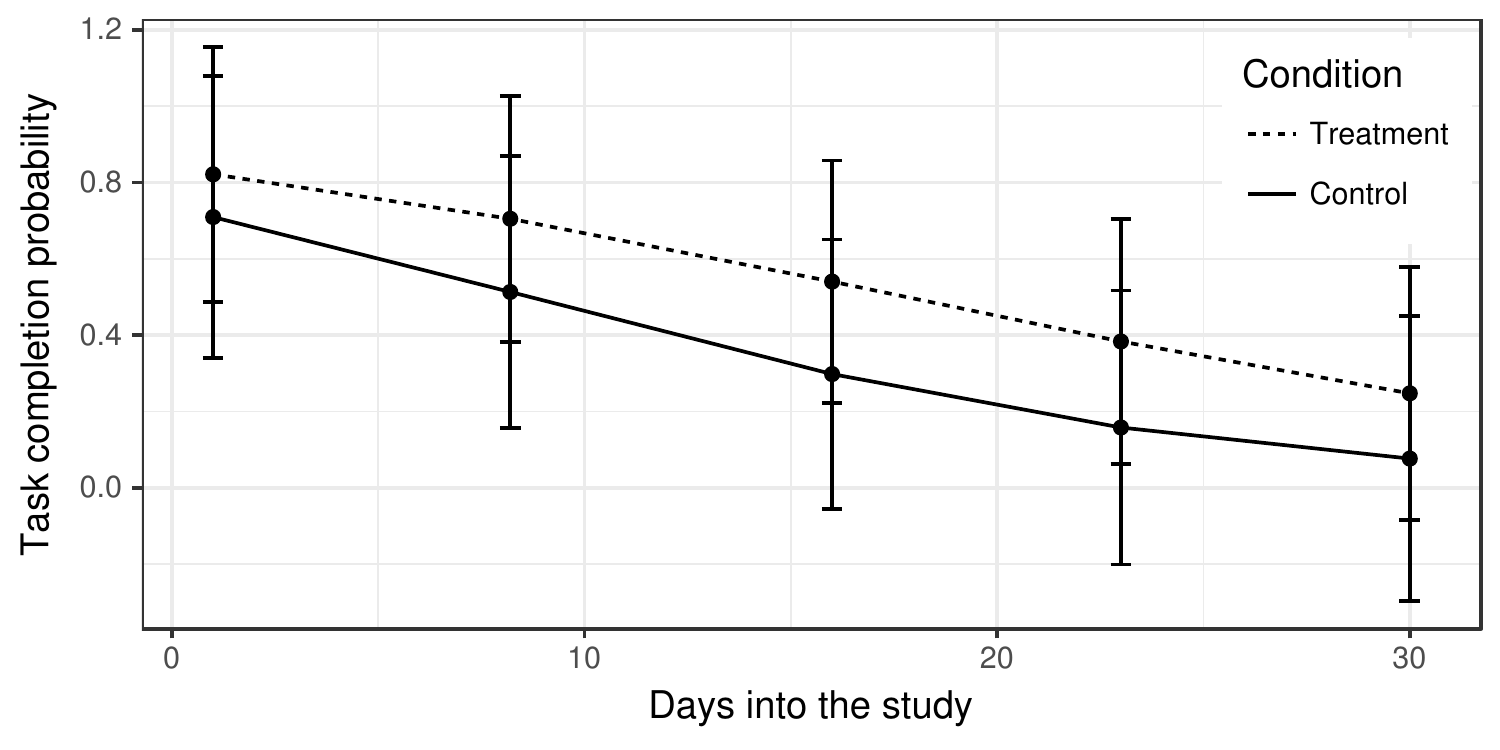}
}
\caption{Interaction effect of interventions and days into the study.}
\label{fig:interaction}
\end{figure}

Next, we evaluate the significance of the decline in compliance over time, as shown in Figure \ref{fig:lines_compliance_days}, by controlling for the number of days since a participant has engaged with the study (Model 2) and its interaction with the experimental condition (Model 3). Figure \ref{fig:interaction} displays the interaction effect of interventions and days into the study on the participants' likelihood of completing the daily tasks. A significant interaction term ($p$<0.05), introduced in Model 3, indicates that the benefits of the treatment condition depend on the number of days a participants spent in the study. As participants spend more time in the study, their compliance tends to decline across the experimental groups. However, treatment participants are more likely to complete the tasks than control participants as the study progresses. That is, a one day increase results in the odds ratios of 0.91305 and 0.88992 for the treatment condition and the control condition, respectively. Note that the odds ratios are obtained from exponentiating the sum of the logistic regression coefficients for \textit{days into the study} and its \textit{treatment interaction term}

In Models 4 and 5, we further assess the main and interaction effects of the inventions by adding control variables for the participants' prior food logging experience (Model 4) as well as their current experience in engaging with the daily tasks (Model 5). After controlling for these factors, the treatment interaction effect remains significant ($p$<0.01) and the results suggest that the effectiveness of our intervention is robust even after adjusting for past experience and self-efficacy. Moreover, the difference in the odds ratios between the experimental conditions increases by 9.5\% in Model 5, a considerable gain from the baseline models. As the study progresses one more day, the odds ratios of the treatment condition and the control condition are 0.92608 and 0.84562, respectively. Unsurprisingly, we find that self-efficacy has a positive effect ($p$<0.01) on the participants' likelihood of completing the tasks, i.e., participants who consider the task to be easy/very easy on a specific day are more likely to finish it and vice versa. Interestingly, past logging experience has no significant effect on daily task completion. This may be explained by the fact that logging food diary is merely a part of the requirements for completing the daily task. Remembering to filling out all daily surveys \textit{in time} is perhaps more demanding to most participants. Thus, having some familiarity with food logging is not an major contributor to the participants' compliance.

Lastly, we examine the main effects of personal characteristics, i.e., gender, age, and grit, and their treatment interaction effects on the participants' day-to-day compliance. These variables are included in Models 6, 7, and 8 as shown in Table \ref{tbl:mixed_effects}. Unlike past studies \cite{Kato-Lin2016}, none of the personal characteristics have any significant impact on the likelihood of completing the daily tasks after adjusting for other factors. Although participants with higher grit scores are more likely to complete the daily tasks than those with lower scores, grit does not significantly impact the participants' compliance, which is surprising. As suggested by Duckworth et al. \cite{Duckworth}, grit constitutes \textit{perseverance} and \textit{passion} for long-term goals. Therefore, one possible explanation is that, even though the participants have voluntarily signed up for our study, they may not have strong passion for the day-to-day self-tracking and reporting tasks, making grit less effective in this context. Overall, the full model (Model 8) shows an encouraging result indicating that the effect of our random-loss incentive on day-to-day compliance is significant even after controlling for all other factors. According to the full model, the odds of completing the daily tasks for treatment participants and control participants when the time of the study increases by 1 day are 0.93723 and 0.87399, respectively.

\subsection{Post-intervention self-tracking behavior}
In this section, we examine the intervention effects on the participants' change in their dietary self-tracking habit before the interventions were administered and after they were removed. Approximately one month after the last participant had completed the study, we collected the food logging data from all participants in the following 30 days after their study period was over. Table \ref{tbl:post_intervention} displays the participants' food logging behavior in the last 30 days before and after the intervention. First, there is a significantly larger number of food loggers during the post-intervention than the pre-intervention phases (40 versus 25; $\chi^2$=3.991, $p$<0.05). The gain is due to changes in the participants' logging habits whereby 25 participants who had never logged any diaries during the pre-intervention phase continued to log their food intake after the study was over, whereas 10 participants who had previously logged food diaries during the pre-intervention phase became inactive thereafter. However, there is no significant difference in the numbers of \textit{active} food loggers (defined as those who logged at least 7 days of food diary in the last 30 days) between the two periods. Amongst the treatment and the control conditions, we found no significant differences in the numbers of food loggers (21 versus 19 and) and active loggers (9 versus 10) post-intervention. Lastly, on average, all participants logged their diaries more frequently during the post-intervention than the pre-intervention phases (2.29 versus 1.3) but the difference is not significant. Within a subset of participants ($N$=92) who logged their diaries during both periods, there is a significant increase in the average food diary days from the pre-intervention to the post-intervention phases (3.47 versus 6.09; $V$=1414, $p$<0.01) but there is no difference between the conditions.

\begin{table}[thbp]
\centering
\caption{Pre/post-intervention logging behavior.}
\label{tbl:post_intervention}
\scalebox{0.75}{
\begin{threeparttable}
\begin{tabular}{@{}llll@{}}
\toprule
Pre-intervention & Treatment & Control & All \\ \midrule
Number of food loggers & 13 & 12 & 25 \\
Number of active food loggers & 5 & 7 & 12 \\
Number of logged diary days (mean) & 1.12 & 1.53 & 1.3 \\ \midrule
Post-intervention &  &  &  \\ \midrule
Number of food loggers & 21 & 19 & 40 \\
Number of active food loggers & 9 & 10 & 19 \\
Number of logged diary days (mean) & 1.84 & 2.84 & 2.29 \\ \midrule
Pre/post-intervention changes &  &  &  \\ \midrule
Number of new food loggers & 12 & 13 & 25 \\
Number of new active food loggers & 2 & 6 & 8 \\ 
Number of existing food loggers who gave up & 4 & 6 & 10 \\ \bottomrule
\end{tabular}
\begin{tablenotes}
\small
\item Active food loggers are defined as users who recorded at least 7 days of food diary in the last 30 days.
\end{tablenotes}
\end{threeparttable}}
\end{table}

\section{Concluding Discussion}
\label{sec:discussion}


To our knowledge, this study is the first randomized trial to demonstrate that framing financial incentives as random loss is effective in increasing individuals' engagement with dietary self-tracking and self-reporting. Particularly, we show that a small unpredictable loss can serve as a fairly powerful motivator to promote compliance and alleviate participants' insensitivity to repeated interventions. The results are encouraging especially since it is well-established that loss-framed incentive is a strong motivator for health behavior change \cite{Patel2016, Volpp2008}. Our results indicate that individuals who were in the random-loss incentive program are significantly more likely to complete the daily tasks than those who were in the fixed-loss incentive program after controlling for time spent in the study and other personal characteristics, such as gender, age, prior self-tracking experience, self-efficacy, and grit. Unlike previous work \cite{Kato-Lin2016}, we found that the heterogeneity of participants did not affect the outcomes of the interventions. For instance, male and female participants did not differ significantly in their likelihood to complete daily tasks. More surprisingly, there is no significant difference in compliance amongst individuals with differing self-tracking experience and grit. This may be explained by the facts that: (1) prior food logging experience marginally contributes to the success of completing self-report surveys, which is arguably a more demanding requirement of the study; and (2) highly gritty participants may not necessarily be more passionate about self-tracking and self-reporting due to its tediousness nature and therefore are not more likely to persistently pursue the goals than less gritty participants. At the post-invention phase, we found no significant difference in the effectiveness of the incentives in sustaining self-tracking behavior for the next 30 days after the interventions were discontinued. Although there was an initial increase in the number of food loggers and self-tracking activities during the first post-intervention week, these numbers substantially declined in the subsequent weeks as more individuals disengaged from dietary self-tracking. The fact that the self-tracking behavior was not sustained post-intervention is not surprising. Unlike typical health intervention studies, participants in our study were not instructed to use self-tracking tools as a means to achieve specific health goals nor were they motivated beyond the primary scope of the study, i.e., contributing their lifestyle data to research about dietary habits. Consistent with recent studies \cite{Finkelstein2016, Patel2016}, our results confirm the importance of a long-term financial incentive program in self-tracking studies.


The results have several implications for future precision medicine and mobile-health (mHealth) studies and interventions. First, our random-loss incentive is proven to be effective in promoting long-term engagement with self-tracking regardless of individuals' personal characteristics, making it suitable for population-level intervention programs. Next, the cost-effectiveness of our incentive implies that improved compliance is attainable with relatively modest cost thanks to the effects of loss aversion and unpredictability. In our setting, the mean cost per participant, including both the cash and the smartwatch rewards, is less than S\$40. Lastly, in terms of data integrity, we observe very little noise in the food diary entries and survey responses contributed by participants in our study. Apart from inaccurate nutritional facts of some food items which we discussed earlier, we find no evidence of misbehavior by our participants. 

\textbf{Limitations:} Our study design specifically focused on dietary self-tracking and self-reporting compliance in healthy individuals and did not consider health outcomes, e.g., improved BMI, which might serve as additional motivation for participants. Understandably, this would have required a different study design and population. Although we initially aimed to representatively recruit participants from Facebook users in Singapore, the majority of participants eventually enrolled were female working adults which may potentially bias the generalizability of the findings. Lastly, future studies should further examine the effects of unpredictability (i.e., the Random Rewards design) on gain-framed and loss-framed incentives in the same settings.



\section{Acknowledgement}
This research is supported by the National Research Foundation, Prime Minister's Office, Singapore under its International Research Centres in Singapore Funding Initiative.

\balance{}

\bibliographystyle{ACM-Reference-Format}
\bibliography{main}

\end{document}